\documentclass{article}

\usepackage{arxiv}

\usepackage[utf8]{inputenc} 
\usepackage[T1]{fontenc}    
\usepackage{hyperref}       
\usepackage{url}            
\usepackage{booktabs}       
\usepackage{amsfonts}       
\usepackage{microtype}      
\usepackage{graphicx}
\usepackage{natbib}
\title{Probabilistic Neural-Network Based 2D Travel Time Tomography}

\date{} 					

\author{
  Stephanie Earp \\
  Department of Geosciences\\
  University of Edinburgh\\
  \texttt{stephanie.earp@ed.ac.uk} \\
   \And
	Andrew Curtis\\
  Department of Geosciences\\
  University of Edinburgh\\
  and \\
    Institute of Geophysics\\
  ETH Zurich\\
    \texttt{andrew.curtis@ed.ac.uk} \\
}

\begin{document}
\maketitle

\begin{abstract}
Travel time tomography for the velocity structure of a medium is a highly non-linear and non-unique inverse problem. Monte Carlo methods are becoming increasingly common choices to provide probabilistic solutions to tomographic problems but those methods are computationally expensive. Neural networks can often be used to solve highly non-linear problems at a much lower computational cost when multiple inversions are needed from similar data types. We present the first method to perform fully non-linear, rapid and probabilistic Bayesian inversion of travel time data for 2D velocity maps using a mixture density network. We compare multiple methods to estimate probability density functions that represent the tomographic solution, using different sets of prior information and different training methodologies. We demonstrate the importance of prior information in such high dimensional inverse problems due to the curse of dimensionality: unrealistically informative prior probability distributions may result in better estimates of the mean velocity structure, however the uncertainties represented in the posterior probability density functions then contain less information than is obtained when using a less informative prior. This is illustrated by the emergence of uncertainty loops in posterior standard deviation maps when inverting travel time data using a less informative prior, which are not observed when using networks trained on prior information that includes (unrealistic) a priori smoothness constraints in the velocity models. We show that after an expensive program of training the networks, repeated high-dimensional, probabilistic tomography is possible on timescales of the order of a second on a standard desktop computer.
\end{abstract}

\section{Introduction}
Seismic travel time tomography is often used to reconstruct images of the interior of the Earth \citep{Aki1977,Dziewonski1987,Mordret2013}, but is a significantly non-linear and non-unique inverse problem. To find solutions with minimal computation, the physics relating local wave speed to measured travel times is usually simplified by linearization \citep{Rawlinson2010}, but this creates large differences between linearized and true probabilistic solutions \citep{Galetti2015}. Increases in compute power now allow fully nonlinear Monte Carlo sampling solutions to be found without linearisation, to solve problems in 2D \citep{Bodin2009,Galetti2015} and 3D \citep{Hawkins2015,Piana2015,Zhang2018,Zhang2019}. Using Bayesian methods, such solutions provide samples (example tomographic models) that fit the data to within their measurement uncertainties, are consistent with available prior information, and are distributed according to the posterior probability density function (pdf) across the parameter space; this pdf constitutes the full solution of tomographic problems. Nevertheless, such solutions are acquired at significant expense, typically requiring weeks of compute time for realistic data sets and expensive storage of large sample sets.

An alternative approach to estimate the posterior pdf is to use prior sampling \citep{Devilee1999,Kaufl2016}. In this case samples are created before inference using only available prior knowledge. The set of samples can then be interrogated for examples that are consistent with any particular data set (a method called \textit{resampling} \citep{Sambridge1999}) or used to parametrise a function that relates data to models which can then be used to solve the inverse or inference problem \citep{Roth1994}.

In this work we use a neural network-based method to perform the inversion. Neural networks (NNs) can approximate any nonlinear relationship between two parameter spaces, given a so-called training set of example pairs of dependent and independent parameter values under that relationship \citep{Bishop1995}. In travel time tomography the forward solution is known and calculable, but the inverse solution is highly non-linear and non-unique. In such cases the forward computation can be used to create the prior set of samples known as a \textit{training set}, of random models drawn from the prior pdf; these can be used to train the neural networks to approximate the inverse mapping. The prior samples are only needed during the training process which needs only to be performed once - thereafter NNs can be evaluated relatively efficiently. This allows the inference step to be run rapidly for any new data set on standard desktop computers, and the overall cost of the method per tomographic problem decreases rapidly with the number of problems to be solved.

Neural network-based inversion methods have been applied to various tomography problems in the past. \cite{Roth1994} first used NNs to estimate subsurface velocity structure from active source seismic waveforms, \cite{Moya2010} performed velocity inversion with a neural network using waveform data from earthquakes, and \cite{Araya-Polo2018} used semblance gathers as input to a network to invert for velocity structure. \cite{Gupta2018} used a convolutional network to learn an ensemble of simpler mappings in a low-dimensional space before reconstructing the image by combining the mappings. Dictionary learning methods \citep{Mairal2014} create sparse representations of the data and can be used to create a set of representations of features. \cite{Bianco2018} performed linearized 2D surface wave travel time tomography using dictionary learning to regularise the inversion.

The methods mentioned above and in \cite{Kong2018} all provide only deterministic solutions to the inversion. Since the solution to tomographic problems is always non-unique, in order to assess the worth of any model estimate we require that neural networks produce full probabilistic information about the set of models in the inverse problem solution (the posterior pdf). \cite{Devilee1999} solved the first probabilistic geophysical inverse problem using NNs. They proposed a variety of methods to train NNs to provide discretised Bayesian posterior pdfs. Mixture density networks (MDNs) are a class of augmented neural networks that output a probability distribution that is defined as a sum of analytic pdf kernels such as Gaussians \citep{Bishop1995}. MDNs can be trained such that for any input data this distribution approximates the posterior pdf. These methods have been used at a global scale to invert surface wave velocities for global crustal thicknesses and seismic velocities \citep{Meier2007a,Meier2007} and for water content in the mantle transition zone \citep{Meier2009}, at a reservoir scale to infer petrophysical parameters from velocities \citep{Shahraeeni2011,Shahraeeni2012}, for earthquake source parameter estimation \citep{Kaufl2014,Kaufl2015} and to assess the uncertainty in model parameters of the Earth's global average (1-dimensional) radial velocity structure from P-wave travel time curves \citep{DeWit2013}. They have also been used in conjunction with Markov random fields and other statistical and graphical models to solve geophysical inverse problems with spatially sophisticated prior information \citep{Nawaz2017,Nawaz2018,Nawaz2019}. They have been used in conjunction with seismic gradiometry to perform near-real time 3D surface wave tomography \citep{Cao2020}. These studies demonstrated that the pdf obtained from an MDN is comparable to a Monte Carlo sampling solution but is obtained at much lower computational cost in the cases where similar inverse problems must be solved repeatedly with different data sets, and that at the moment of application MDNs provide probabilistic solutions almost instantaneously.

We show for the first time that MDNs can perform fully non-linear, rapid and probabilistic 2D tomography from travel time data. We compare different methods for creating the prior training set and performing the neural network inversion. The networks create approximate mean velocity models and estimates of the full marginal posterior pdf's, virtually instantaneously. Thus, in return for accepting approximate posterior pdfs we obtain a significant computational saving compared to Monte Carlo methods.
\section{METHOD}\label{method}
\subsection{Bayesian Inference}\label{Bayes_Inf}
We wish to solve tomographic inverse problems in a probabilistic framework to find the posterior distribution of velocity models $\textbf{m}$ that fit some given data $\mathbf{d}$, written as $p(\textbf{m} \mid \textbf{d})$. This is defined as \citep{Tarantola2005}:
\begin{equation}\label{Bayes}
p( \textbf{m}\mid \textbf{d}) = k\, p(\textbf{d} \mid \textbf{m}) \, p(\textbf{m})
\end{equation}
where $p(\textbf{m})$ represents the prior probability density on the model space, $p(\textbf{d} \mid \textbf{m})$ represents the conditional probability of some data given the model (known as the likelihood) and $k$ is a normalisation constant. In multidimensional problems, where the dimensionality of $\textbf{m}$ is greater than 1, we often need to make inferences about a single parameter with index $i$ and hence must calculate the marginal posterior distribution $p( m^i\mid \textbf{d})$. This can be obtained by integrating over all parameters $j$ that are not of interest:
\begin{equation}\label{marg_post}
p( m^i\mid \textbf{d}) = \int_{\textbf{m}_j\neq \textbf{m}_i} p(\textbf{m} \mid \textbf{d})\, d\textbf{m}_j
\end{equation}
In this study we focus on estimating marginal distributions $p( m^i\mid \textbf{d})$, and posterior trade-offs between pairs of individual parameters.
\subsection{Mixture Density Networks}\label{MDNs}
Neural networks are essentially mathematical mappings that emulate the relationship between two parameter spaces. Given a set of $N$ data-model pairs $\{(\mathbf{d}_i,\mathbf{m}_i): i =1,...,N\}$, where $\mathbf{m}_i$ is the model used to generate the data $\mathbf{d}_i$ under some forward relation, NNs can be trained to model an arbitrary non-linear inverse function from $\mathbf{d}$ to some properties of the set of models $\mathbf{m}$. In this paper we use a class of neural networks called mixture density networks, that can be trained to output the probability of any model $\mathbf{m}$ given some fixed (measured) data $\mathbf{d}$, written as $p(\textbf{m} \mid \textbf{d})$. The probability distribution is approximated using a sum (called a mixture) of Gaussians:
\begin{equation}\label{MDN_mix}
p(\textbf{m} \mid \textbf{d}) \simeq \sum_{i=1}^{M}\alpha_i(\textbf{d})\Theta_i(\textbf{m}|\textbf{d})
\end{equation}
where  $\alpha_i$ is called the mixture parameter that attaches relative importance to each Gaussian kernel, $M$ is the number of Gaussians in the mixture, and $\Theta_i$ are here defined to be Gaussian kernels with a diagonal covariance matrix given by
\begin{equation}\label{Gauss_mix}
\Theta_i(\textbf{m} \mid \textbf{d}) = \frac{1}{\prod^c_{k=1}(\sqrt{2\pi}\sigma_{ik}(\mathbf{d}))} exp\left \{-\frac{1}{2}\sum_{k=1}^c\frac{(\mathbf{m}_i-\mu_{ik}(\mathbf{d}))^2}{\sigma^2_{ik}(\mathbf{d})} \right \}
\end{equation}
where $c$ is the dimensionality of $\textbf{m}$, $\mu_{ik}$ is the $k$th element of the $i$th kernel in the mixture, $\sigma_{ik}$ is the standard deviation of the $k$th diagonal element of the $i$th kernel in the mixture, and both $\mu_{ik}$ and $\sigma_{ik}$ are outputs of a trained NN. The network is trained by minimising the negative log likelihood of the pdf in Equation \ref{Gauss_mix}, equivalent to maximizing the likelihood of the pdf \citep{Bishop1995}. For a more comprehensive general introduction to MDNs we refer the reader to \cite{Bishop1995}, or to \cite{Meier2007a} and \cite{Shahraeeni2011} for detailed descriptions with applications in geophysics.

Network training is performed using gradient-based optimization of the network's internal parameters. The particular trained NN obtained is therefore sensitive to the random parameter initialization and to the network configuration (internal structure). We train an ensemble of multiple networks with different configurations and combine them to give a group of networks - a so-called \textit{mixture of experts}. In theory networks trained independently may make good predictions for different reasons and under different inputs (in our case, data vectors); using a combination of networks therefore often results in better generalisation of performance to unseen data and improves prediction accuracy \citep{Dietterich2000}. We construct the ensemble by a weighted average of network outputs, where each weight is determined by the performance of the associated network on the test data set (or simply \textit{test set}). The posterior probability distribution is thus estimated by 
\begin{equation}\label{ensemble}
p(\textbf{m} \mid \textbf{d}) \simeq \sum_{i=1}^{M} \sum_{j=1}^{c} \frac{E_{i}\alpha_{ij}}{\sum^{M}_{k=1} E_k}(\textbf{d})\Theta_{ij}(\textbf{m}|\textbf{d})
\end{equation}
where $E_{i}$ is the negative exponential of the error on the test dataset of the $i$th kernel. The final estimate of probability distribution $p(\textbf{m} \mid \textbf{d})$ contains $cM$ Gaussian kernels.
\subsection{Model Parametrisation and Traveltime Data}\label{mod_data}
\begin{figure}
\centering
\includegraphics[width=0.6\textwidth]{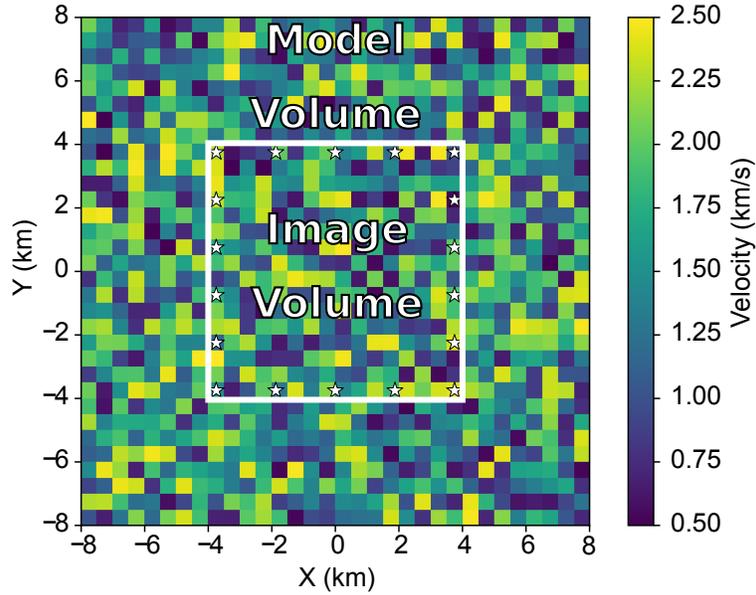}
\caption{Geometry of velocity models. Larger model with limits (-8,8) in the X and Y direction is the \textit{Model Volume} within which the travel-times are calculated. The smaller model bounded by a white box with limits (-4,4) in the X and Y direction is the \textit{Image Volume} which we wish to image. White stars represent the location of co-located sources and receivers, between which travel time data are obtained.}
\label{fig:models_full}
\end{figure}

\begin{figure}
\centering
\includegraphics[width=\textwidth]{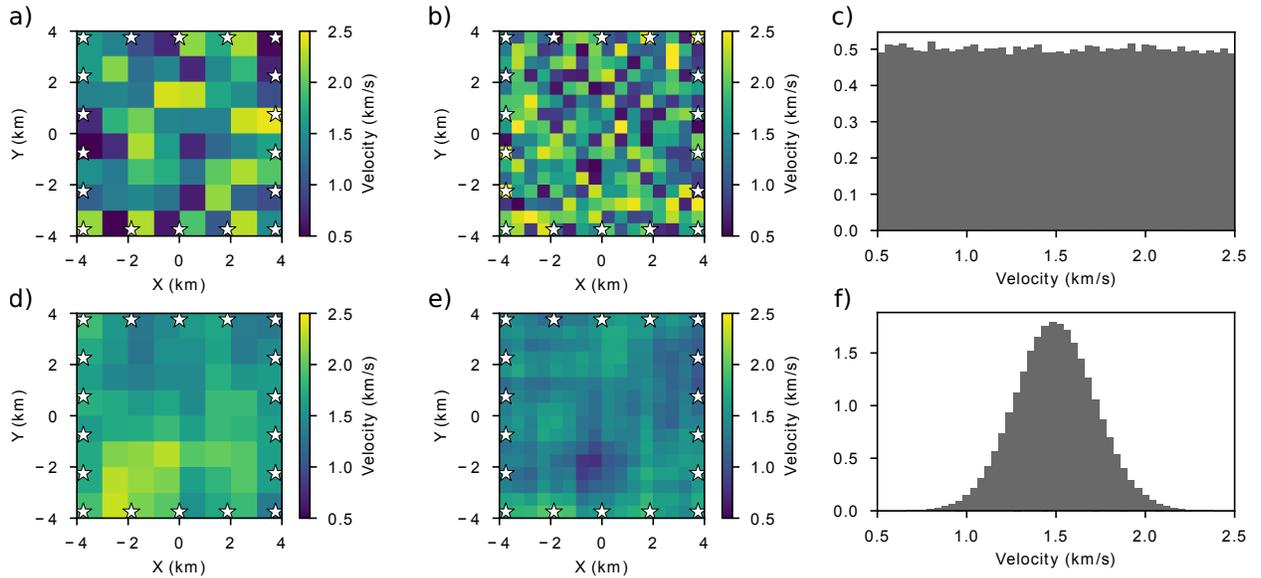}
\caption{Example velocity models from the 4 training sets that are randomly selected from Uniform distributions on an (a) 8 by 8 grid and (b) 16 by 16 grid, or are randomly selected and then smoothed with a spatial averaging filter on a (d) 8 by 8 grid and (e) 16 by 16 grid. White stars represent the location of co-located sources and receivers. The prior distribution of the training set is shown for one cell in the model given a fixed neighbouring cell  for (c) models selected from a Uniform random distribution and (f) similar models after spatial smoothing.}
\label{fig:models}
\end{figure}

\begin{figure}
\centering
\includegraphics[width=0.7\textwidth]{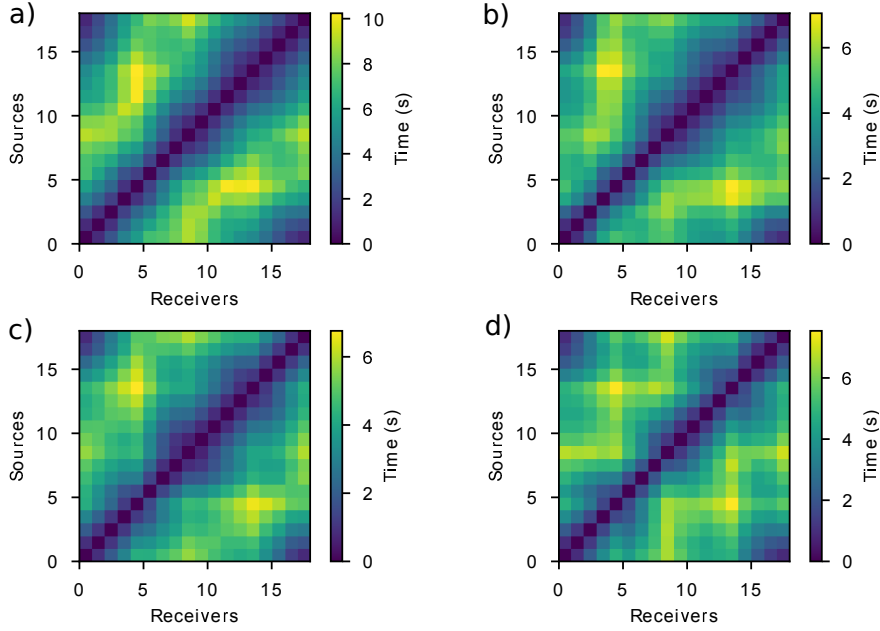}
\caption{Corresponding data from the four velocity models in Figure \ref{fig:models} that are randomly selected from Uniform distributions on an (a) 8 by 8 grid and (b) 16 by 16 grid, or are randomly selected and then smoothed with an averaging filter on a (c) 8 by 8 grid and (d) 16 by 16 grid.}
\label{fig:data}
\end{figure}

We define the geometry of our tomography problem to be that shown in Figure \ref{fig:models_full}. We fix the locations of 18 wave energy sources and receivers (shown in Figure \ref{fig:models}), and parametrise the wave speed or velocity across the \textit{Model Volume} within which the forward relationship predicts travel times of the first arriving energy between any source-receiver pair. Travel times $\textbf{d}_i$ between all possible source-receiver pairs are calculated using an eikonal raytracer \citep{Rawlinson2004,Rawlinson2005}. The traveltimes from the 4 velocity models shown in Figure \ref{fig:models} are shown in Figure \ref{fig:data}. Such travel times are used herein to image the velocity structure within the smaller \textit{Image Volume} - wave speeds outside of that area are disregarded and thus constitute nuisance parameters. We use a larger volume to calculate the forward relationship to avoid raypaths travelling along the boundary of the model and causing misleading travel times.

We construct four separate training sets, each of 2.5 million discretised models where each model represents a 2D heterogeneous velocity structure. Two of these training sets are created on an 8 x 8 coarser grid of cells and two are created on a 16 x 16 finer grid of cells within the \textit{Image Volume} (and the same resolution extends throughout the \textit{Model Volume}). Each of the four datasets is created by selecting a random wave speed in each cell independently from the Uniform prior distribution $U(0.5km/s,2.5km/s)$. All models in one finer data set and one coarser data set are then smoothed using a 2D averaging filter window which was square of size 5x5 cells for the finer model and 3x3 cells for the coarser model. Thereafter the velocities are normalised to the same absolute range as the original random models for ease of comparison of results. Then, the travel times between all source-receiver pairs are calculated for all models, in all four training sets (examples are given in Figure \ref{fig:data}).

With this method we create training sets with two different amounts and types of prior information. The two sets of random unsmoothed velocity models have relatively weak prior information with no correlations between neighbouring cells. This has the advantage that any type of velocity contrast between neighbouring cells would be consistent with the prior pdf and hence can in principle be imaged using the associate trained network given sufficiently informative data (see below). This is demonstrated by the uniform distribution of the histogram in Figure \ref{fig:models}c which shows the probability of the velocity of the adjacent cell given that the velocity of the central cell is 1.5km/s. On the other hand this implies that the prior pdf is Uniform over a 64- and 256- dimensional space for the coarser and finer training sets respectively; these spaces are therefore extremely sparsely sampled by the 2,500,000 training set models due to the curse of dimensionality \citep{Curtis2001}. This implies that over most of these two spaces the prior pdf is entirely unrepresented by `proximal' samples.

The two sets of smoothed velocity models embody stronger prior information as the speeds in neighbouring cells are correlated. This is demonstrated in Figure \ref{fig:models}f where the distribution of possible velocities in adjacent cells given that the velocity of the central cell is 1.5km/s is approximately Gaussian. This means that models with larger velocity contrasts between neighbouring cells are not represented in the training data set and hence will be precluded from inversion results. This may or may not be advantageous depending on the true prior information about the form of the structure being imaged. However, it has the advantage that the effective space (manifold) of models consistent with the prior information is considerably smaller than that for the smoothed models, so that the finite-sized training set may better represent the form of the prior pdf.
\section{RESULTS}\label{results}
\subsection{Network Configurations}\label{networks}
We train separate MDNs to predict the marginal probability distribution $p( m^i\mid \textbf{d})$ of velocity $m^i$ in cell $i$ in each of the two sizes of models. For the finer datasets we train 4 MDNs and for the coarser datasets we train 8 MDNs at each location $i$. We use different configurations as well as randomly initialised internal network parameters (commonly referred to as weights and biases) for each network because diversity in the ensemble generally leads to better predictions \citep{Dietterich2000}. Appendix A outlines the different network configurations. For each network we use a Gaussian mixture consisting of 15 kernels. The precise number of kernels is not important as long as it is larger than the number required to represent the marginal posterior pdf in each model cell. The network can either reduce the amplitude of the mixture parameter $\alpha_{ij}$ to close to zero to remove unnecessary kernels, or can combine unnecessary kernels by giving them a similar $\mu$ and $\sigma$ to other kernels \citep{Bishop1995}. In practice we found the maximum number of kernels with significant weight used in any mixture was 8.

We also train networks to invert for the full model (velocities in all cells at once) using a single network. In this case we use a convolutional network with 3 convolutional layers followed by 3 fully connected layers and 15 kernels for the Gaussian mixture. We train 10 networks with 5 different network configurations (each configuration is trained twice with random weight initialisation). Layer sizes were selected using the python library hyperopt \citep{Bergstra2015} and Appendix A gives further description of the networks used. The same network configurations were trained on all four training sets.

For every training run for each network configuration we use 85\% of the training dataset to train the network, 10\% of the dataset as a validation set during training, and 5\% as a test set to evaluate the final network once training has finished. The training set is used in the optimisation of network parameters. The parameters are updated iteratively so that the output of the network best represents the training set sample distribution. To avoid over-fitting the network to the data the cost function is also periodically evaluated over the validation set; when the error on the validation set stops decreasing we end the training optimisation. Once all of the networks have been trained we evaluate the final network performance using the test set and sum the networks across the ensemble using equation \ref{ensemble}.
\subsection{Result Evaluation}\label{res_eval}
We tested our trained networks by applying them to synthetic data sets calculated for velocity models created specifically to test the performance of each type of network. The quality of the mean of the inverted probability distributions of 2D velocity models (comprising 1D marginal posterior pdfs in each model cell in the cases where networks were trained for each cell individually) are compared against the true velocity model using the structural similarity index metric (SSIM). This metric is based on 3 relatively independent comparison measurements: luminance, contrast and structure (Appendix B). SSIM can assume values between -1 and 1: a value of 1 indicates the images are identical, 0 indicates no structural similarity and negative values occur when local structure is inverted. SSIM differs from other quality indicators such as mean squared error (MSE) in that it measures the quality of an image in structure and pixel value compared to a ground truth, rather than the absolute squared errors (which often do not mean much to someone who is trying to interpret the resulting images).

We compare the information gain between the prior $p(\textbf{m})$ and the posterior $p(\textbf{m}| \mathbf{d})$ distribution using the Kullback-Leibler (KL) divergence
\begin{equation}
D_{\mathrm{KL}}\left (p(\textbf{m}| \mathbf{d}), p(\textbf{m}) \right )= \int_{-\infty}^{\infty}p(\textbf{m}| \textbf{d})ln\left (\frac{p(\textbf{m}| \textbf{d})}{p(\textbf{m})}\right )dx
\end{equation}
where a higher $D_{\mathrm{KL}}$ indicates that the posterior pdf has gained information over the prior and $D_{\mathrm{KL}}=0$ occurs when the two distributions are the same. This can be used as an indication of the effectiveness of the network: if $D_{\mathrm{KL}}$ is close to $0$ then the network has been able to learn little, if anything at all, from the data.
\subsection{Prior}\label{prior}
\begin{figure}
\centering
\includegraphics[width=0.8\textwidth]{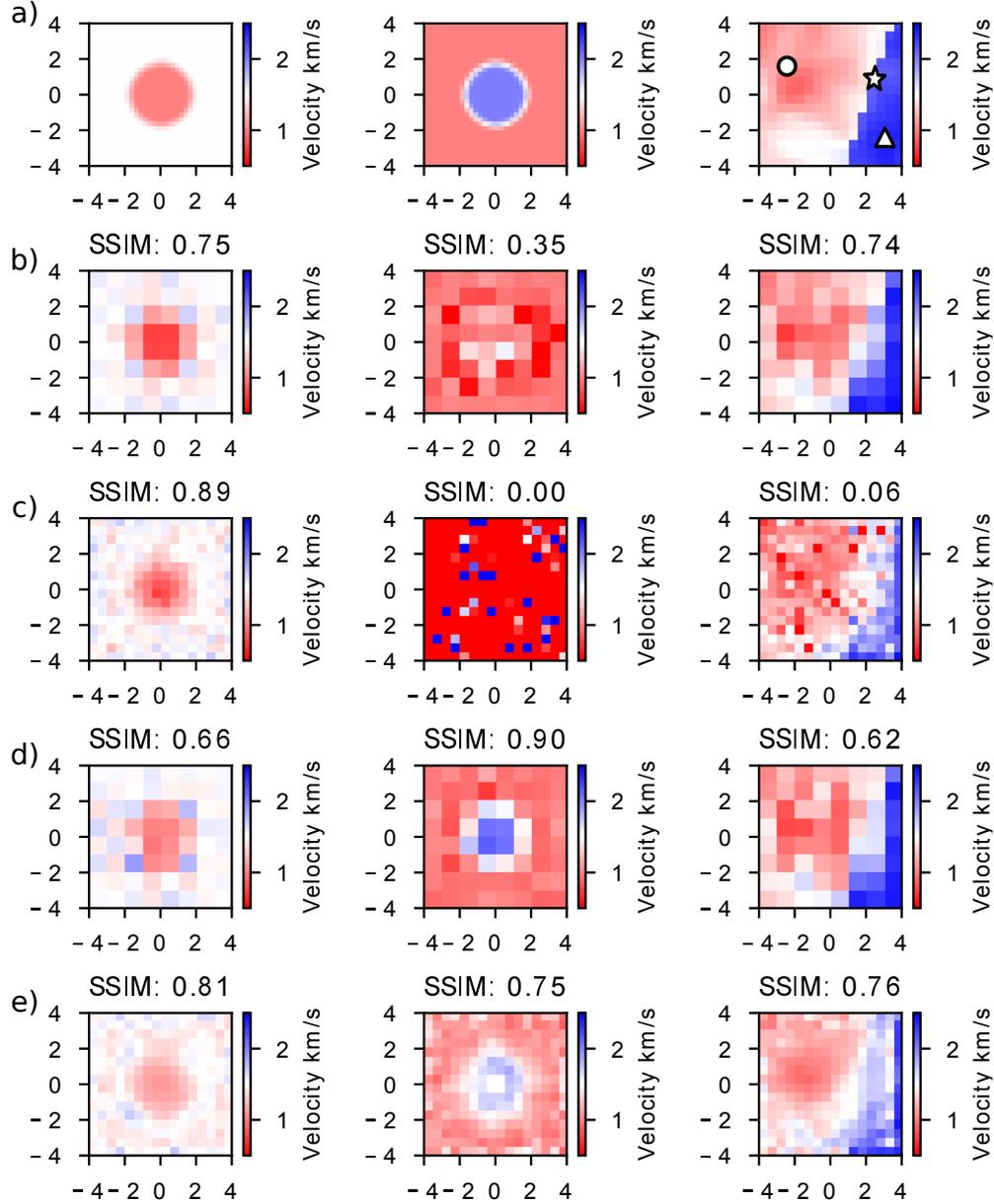}
\caption{(a) True velocity models. Using a randomly generated training set drawn from a Uniform distribution, mean velocities from separate-cell MDN inversions for (b) an 8 x 8 model and (c) a 16 x 16 model, and from full-model MDN inversions for (d) an 8 x 8 model and (e) a 16 x 16 model. The corresponding SSIM values are shown above each result (see Appendix B for definition of SSIM).}
\label{fig:model_rand}
\end{figure}

\begin{figure}
\centering
\includegraphics[width=0.8\textwidth]{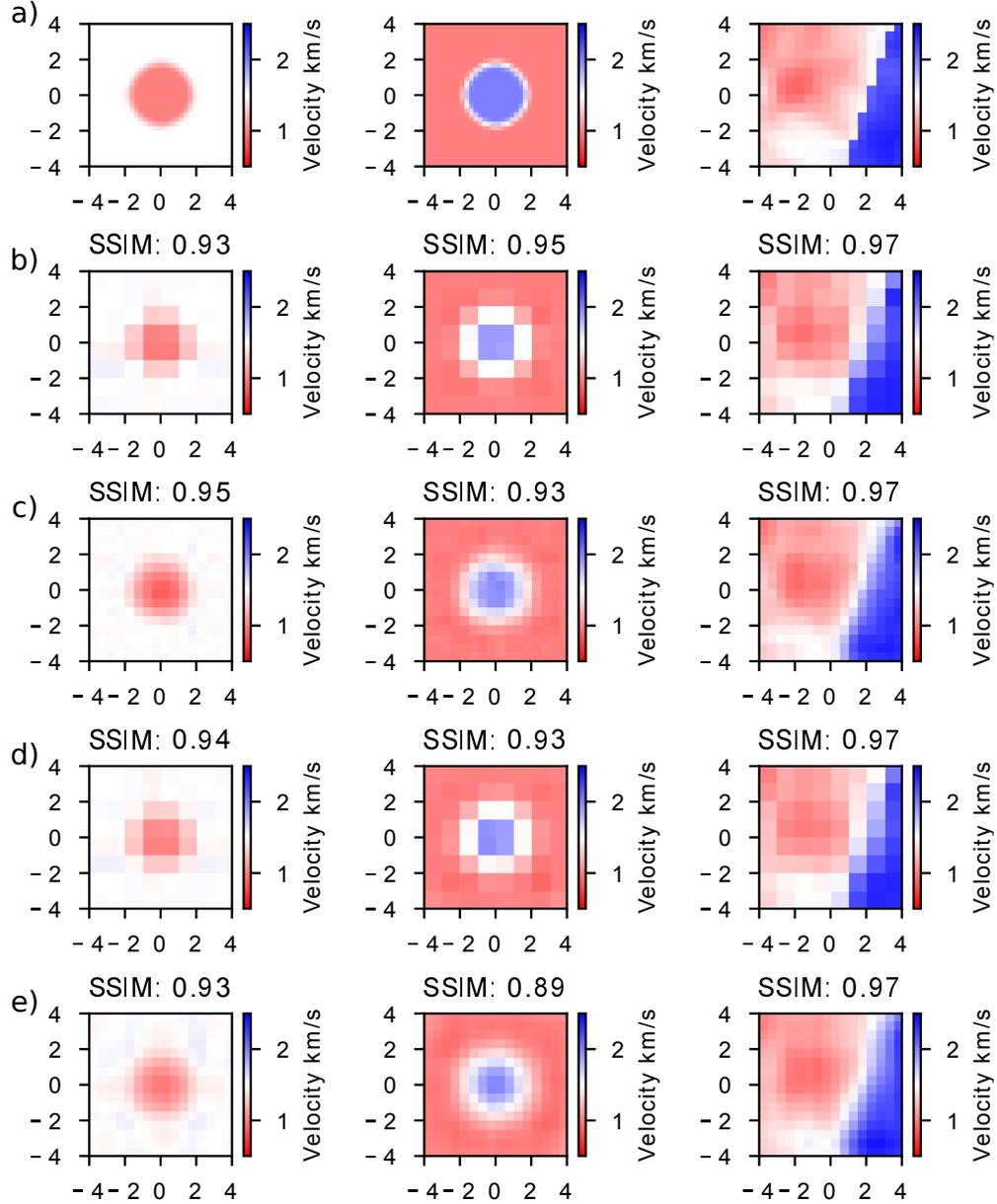}
\caption{(a) True velocity models. Using a training set with spatially smoothed velocities, mean velocities from separate-cell MDN inversions for (b) an 8 x 8 model and (c) a 16 x 16 model, and from full-model MDN inversions for (d) an 8 x 8 model and (e) a 16 x 16 model. The corresponding SSIM values are shown above each result (see Appendix B for definition of SSIM).}
\label{fig:model_smo}
\end{figure}

To show the effect of the prior on our models we inverted synthetic data for the three velocity models shown in Figures \ref{fig:model_rand}a and \ref{fig:model_smo}a using networks trained with weak prior information (unsmoothed training models) in Figure \ref{fig:model_rand} and those trained with stronger prior information (smoothed models) in Figure \ref{fig:model_smo}. The test models were defined on a grid finer than our training sets on a 32x32 grid, which is finer than either of our training sets; this ensures that we evaluate the networks using models that are outside the range of those used for training. For all test models it is clear that with stronger prior information the networks better resolve the velocity structure, shown generally by the much higher SSIM values in Figure \ref{fig:model_smo}b and \ref{fig:model_smo}c compared with the corresponding values in Figure \ref{fig:model_rand}b and \ref{fig:model_rand}c. This is true even though the test models contradict the stronger prior information: they all contain structures that are not smooth.

The velocity model in the left-hand column has a background velocity (cells surrounding the central anomaly) equal to the mean of the prior pdf and a circular low velocity, and is estimated well in both inversions using weaker prior information training sets (Figure \ref{fig:model_rand}). However, even a small increase in complexity in velocity models gives poor inversion results as shown by the central column of velocity models. For these, all the velocities are increased compared to the left column, and in particular the background velocity is increased away from the mean of the prior. In this case the networks with weaker prior information are unable to recover much, if any, of the true structure. If stronger prior information is included in the training set the networks accurately predict a larger variety of velocity models. The true structures of the two circular models in Figure \ref{fig:model_smo} are closely reproduced in the inversion. Sharp contrasts in velocity in the true model are translated to more gradual changes in velocity in the estimates (for both grid sizes) due to the smoothness in the prior pdf. Despite this, the SSIM values show that results are very well correlated with the true model. For the more geologically reasonable model in the right column of Figure \ref{fig:model_smo} which includes a structure that might be generated by a fault, networks trained using stronger prior information on both grid sizes produce models that are nearly identical to the true model. Even though the true model contains a sharp contrast boundary, the inverted models still contain a (slightly smoother) version and the overall structure of the true image is maintained.

\begin{figure}
\centering
\includegraphics[width=0.9\textwidth]{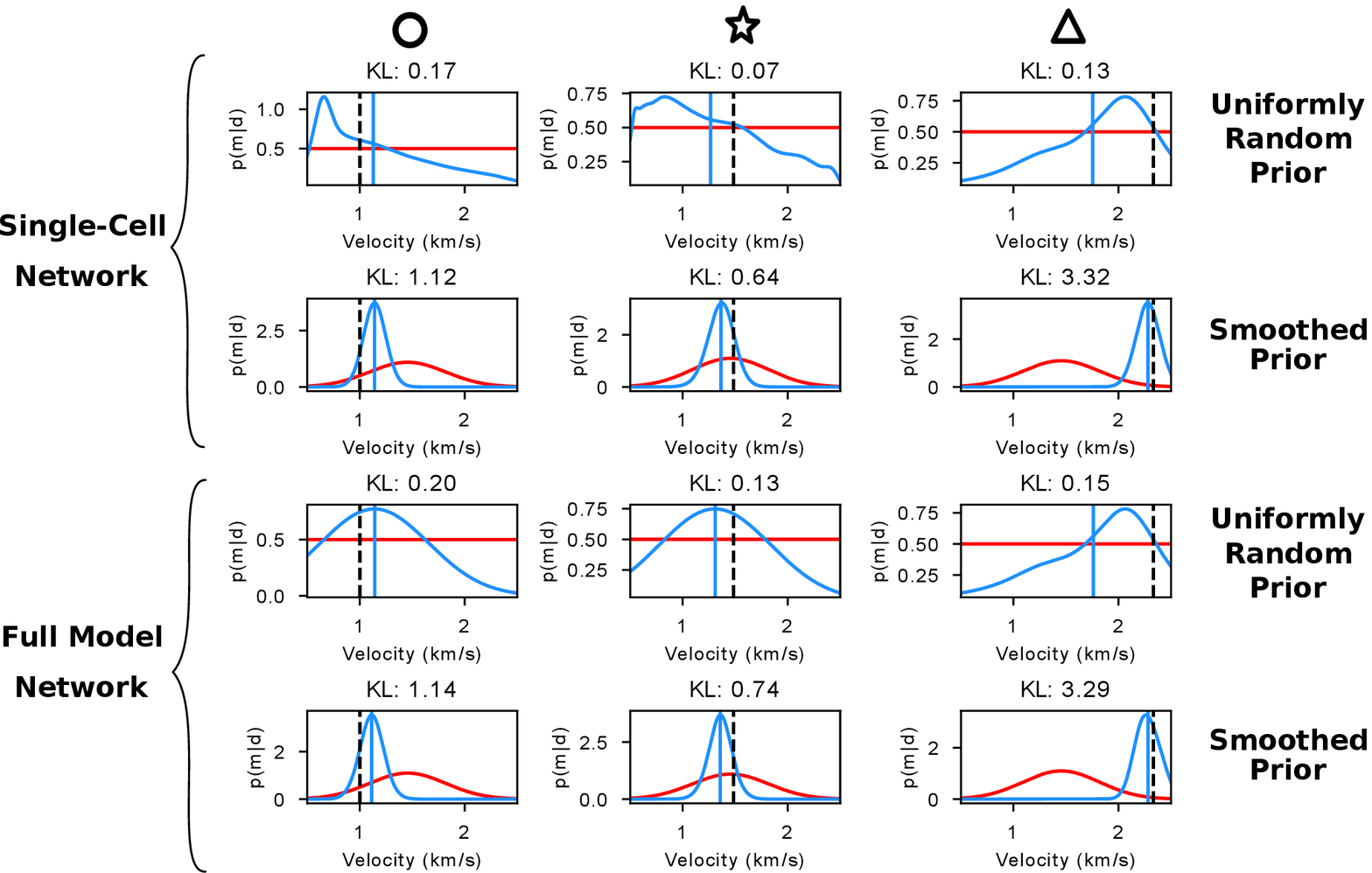}
\caption{Posterior pdfs (blue curves) compared to the prior pdfs (red curves) for the 16 x 16 grid models for three locations shown in the top-right model of Figure \ref{fig:model_rand}: circle (left), star (middle), triangle (right). The rows show results from: (row 1) Separate-cell MDN's using Uniformly random training dataset. (row 2) Separate-cell MDN's using the smoothed training dataset. (row 3) Full-model MDN using Uniformly random training dataset. (row 4) Full-model MDN using the smoothed training dataset. The mean of the posterior is shown by the blue solid line and the true velocity value by a black dashed line. Corresponding KL divergence values are shown above each result.}
\label{fig:pdfs_geol1}
\end{figure}

The effect of stronger prior information is shown in the posterior pdfs in Figure \ref{fig:pdfs_geol1}. We display the posterior marginal pdfs at three locations indicated in the upper right hand model in Figure \ref{fig:model_rand}a: a location in the high velocity zone (triangle), the low velocity zone (circle), and at the edge of the sharp contrast where the inversion struggles to image correctly (star). The KL divergence values are shown above the corresponding posterior marginal pdf. The most striking feature is the much higher KL values for the networks trained with the stronger prior information (rows b and d) indicating a larger information gain in the posterior pdf compared to the prior pdf than is obtained when training with Uniformly random models. In fact, the low KL values for the latter cases imply that nearly no information was gained from the data, and even though a rough approximation of the mean can be found the uncertainties on those values remain large.
\subsection{Model Resolution}\label{res}
Our networks are trained on two sizes of grid cell, a coarser 8 x 8 grid and a finer 16 x 16 grid. Figures \ref{fig:model_rand} and \ref{fig:model_smo} show the results for varying grid size. Training on the finer grid induces a factor of 4 more parameters to estimate from the same data. This means that a larger training set size would be needed to sample the increase in image dimensionality. It would be impossible to sample densely the 256-dimensional space spanned by a 16 x 16 grid, but as our examples show, the networks are still able to invert for some basic structural information (Figure \ref{fig:model_rand}c). When we train our networks with a stronger prior pdf we reduce the effective dimensionality of our problem by introducing a relationship between neighbouring pixels: essentially all prior models and hence most posterior models lie on a significantly lower dimensional manifold that is embedded with the 64- or 256- dimensional spaces. In that case we can obtain reasonable estimates of the true velocity models regardless of grid size (Figure \ref{fig:model_smo}c). 
\subsection{Type of network}\label{type}
For each of the four training sets we trained networks in two different ways. First we trained separate networks to estimate marginal pdf's in each cell so that each network has fewer parameters ($\alpha_{ij}$, $\mu_{ij}$, $\sigma_{ij}$) to estimate. Note that this does not reduce the dimensionality of the overall problem as each velocity cell in the model contributes to the travel time values, and the velocity in any cell depends on the cells surrounding it even if we do not directly invert for them within the same network. It is important to remember that in this case we do not obtain explicit information about trade-offs between neighbouring cells. Those trade-offs are already integrated into the marginal pdf's in equation \ref{marg_post}.

\begin{figure}
\centering
\includegraphics[width=0.7\textwidth]{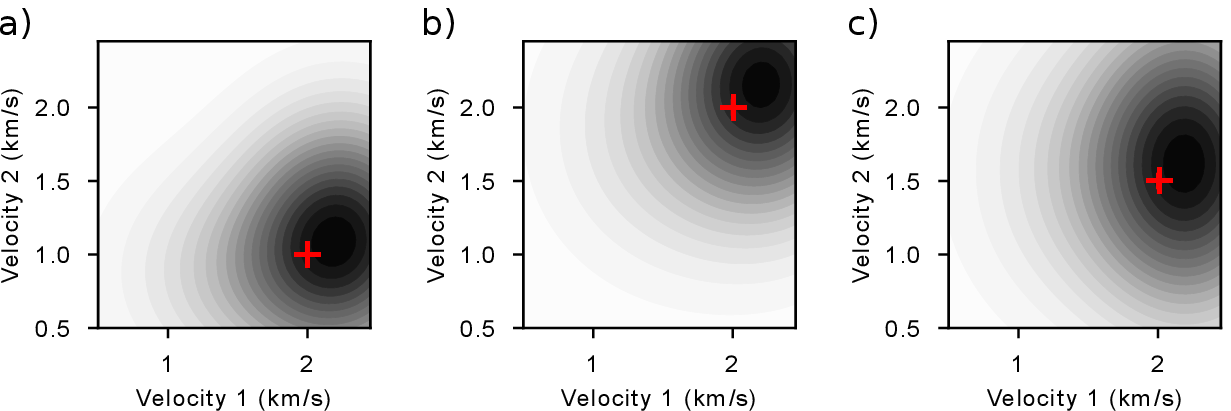}
\caption{Joint pdfs comparing the a pixel inside the velocity high of the central model in Figure \ref{fig:model_rand}a. Velocity 1 is the velocity of a cell in the centre of the velocity high. Velocity 2 is the velocity of a cell a) in the background velocity, (b) at the centre of the velocity high (not the same cell as Velocity 1) and (c) at the edge of the velocity anomaly. }
\label{fig:joint_pdfs}
\end{figure}

We also trained networks to invert for slowness in every cell of the model at once. This increases the number of parameters that the network must estimate but as a result the trade-off between velocity values in adjacent cells can be explored. Examples of the joint marginal pdfs from the central model in Figure \ref{fig:model_rand}a are shown in Figure \ref{fig:joint_pdfs}: the 2D pdfs show few signs of non-linearity, and virtually no indication of the trade-offs that one would expect between velocities in neighbouring cells. This indicates that the results of these networks are unlikely to provide reliable uncertainties.

For models on a coarser grid (Figures \ref{fig:model_rand} and \ref{fig:model_smo} rows b and d), networks perform similarly when using the single cell networks or the full model networks. For models trained on a finer grid, the full model networks perform significantly better than the single cell network as shown in Figure \ref{fig:model_rand}. This is almost certainly because the dimensionality of the problem when training single-cell networks is too large, but by giving the network information about the velocities in neighbouring cells it can better resolve the velocities. This difference is less noticeable when using stronger prior information (Figure \ref{fig:model_smo}b and \ref{fig:model_smo}d).
\subsection{Uncertainty Loops}\label{loops}
A key problem in the field of nonlinear inversion is that there are no standard solutions to which estimated posterior pdf's can be compared in order to verify their quality. In almost all papers that use synthetic tests to assess competing methodologies in high-dimensional problems, the main criterion applied is whether the mean or maximum-likelihood model fits the real (true synthetic) model that was used to generate the synthetic data. This provides no test at all on the rest of the pdf and indeed there is no reason why the mean \textit{should} match the true model in unknown problems - the mean may even be a zero-probability solution (one precluded by the data) \citep{Tarantola2005}. The maximum likelihood (or maximum posterior probability) model is an alternative, but usually an extremely volatile statistic of pdf solutions since those solutions are necessarily formed by focusing across the whole pdf rather than simply on its modes. We therefore require some independent property of posterior pdfs, the existence of which we can use to assess their veracity.

Loops or halos of high uncertainty have been shown to exist in solutions to all travel time tomography problems around anomalies with a spatially sharp and strong contrast in velocity compared to their surroundings \citep{Galetti2015}. Uncertainty loops exist due to non-linear aspects of wave physics and represent uncertainty in the \textit{shape} of such anomalies. They are observed most clearly in fully non-linearized tomographic inversion problems in which rays, velocities and travel times are all varied in concert for each sample considered. We can therefore use the existence of loops in posterior uncertainty as a criterion to check their quality in models with strong and spatially sharp contrasts.

\begin{figure}
\centering
\includegraphics[width=\textwidth]{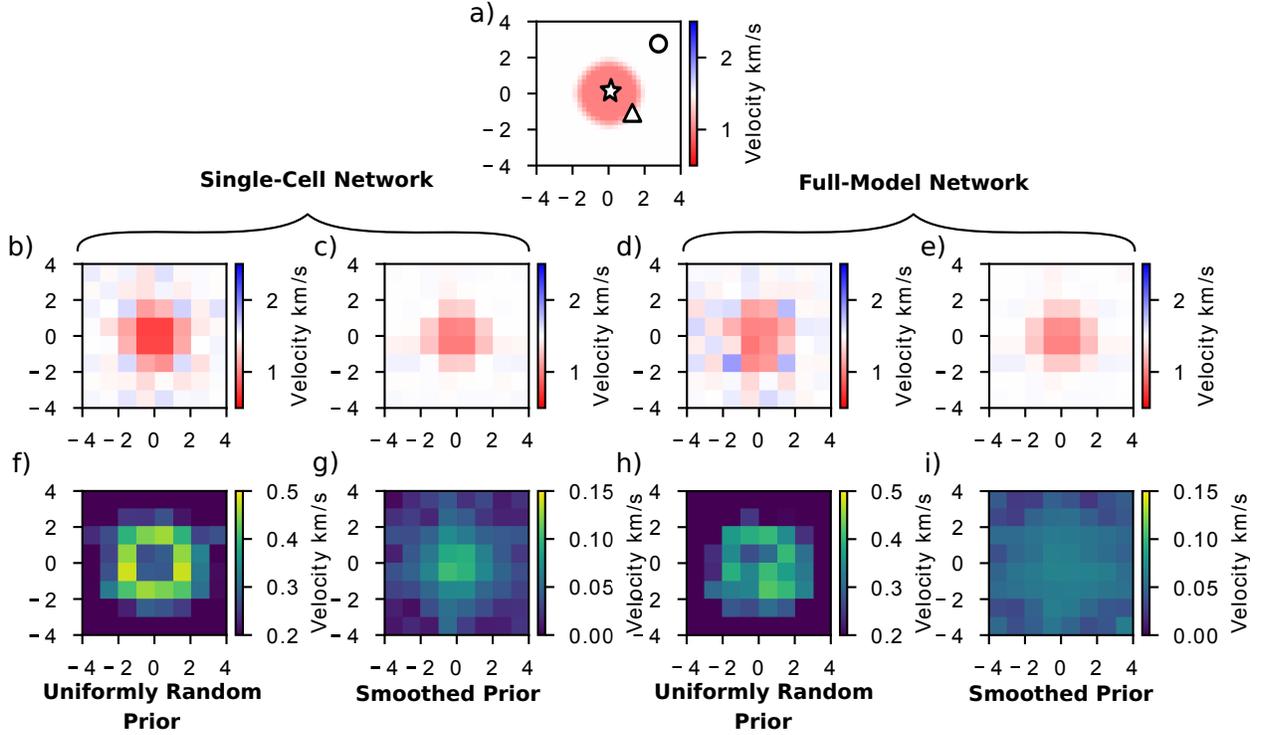}
\caption{(a) True velocity model. For a separate-cell MDN, using a training set from a Uniformly random distribution, results shown are (b) mean velocities and (f) corresponding standard deviations. Using the same type of network with a training set of spatially smoothed velocities we obtain (c) mean velocities and (g) standard deviations. For a full-model MDN, using a training set from a Uniformly random distribution we obtain (d) mean velocities and (h) standard deviations. Using the same type of network with a training set of smoothed velocities we obtain e) mean velocities and (i) standard deviations.}
\label{fig:model_uncer}
\end{figure}

Figure \ref{fig:model_uncer} shows the standard deviations (bottom row) for the results of networks trained on an 8 x 8 grid. Only the networks trained using the training set of Uniformly random velocities (Figure \ref{fig:model_uncer}f and h) exhibit signs of an uncertainty loop. We include the mean (middle row) for comparison of the shape of the velocity anomaly to the loop that surrounds it. The difference between the two priors is clear when comparing Figures \ref{fig:model_uncer}f, \ref{fig:model_uncer}g and \ref{fig:model_uncer}h: for a smoothed prior (Figure \ref{fig:model_uncer}g) the maximum uncertainty is predicted to be in the centre of the anomaly as opposed to the other two images where the uncertainty is lowest at the centre of the anomaly and highest on the margins as expected. However, when inverting for the full-model in a single network (Figure \ref{fig:model_uncer}h) the loop is not as well defined as in Figure \ref{fig:model_uncer}f. Together with the lack of clear trade-off relations in Figure \ref{fig:joint_pdfs} this is evidence that the full-model inversions are less robust than single-cell inversions: as the networks invert for many more parameters at once, they appear not to have been trained so as to fully represent the correct physics of the tomography problem. 

\begin{figure}
\centering
\includegraphics[width=\textwidth]{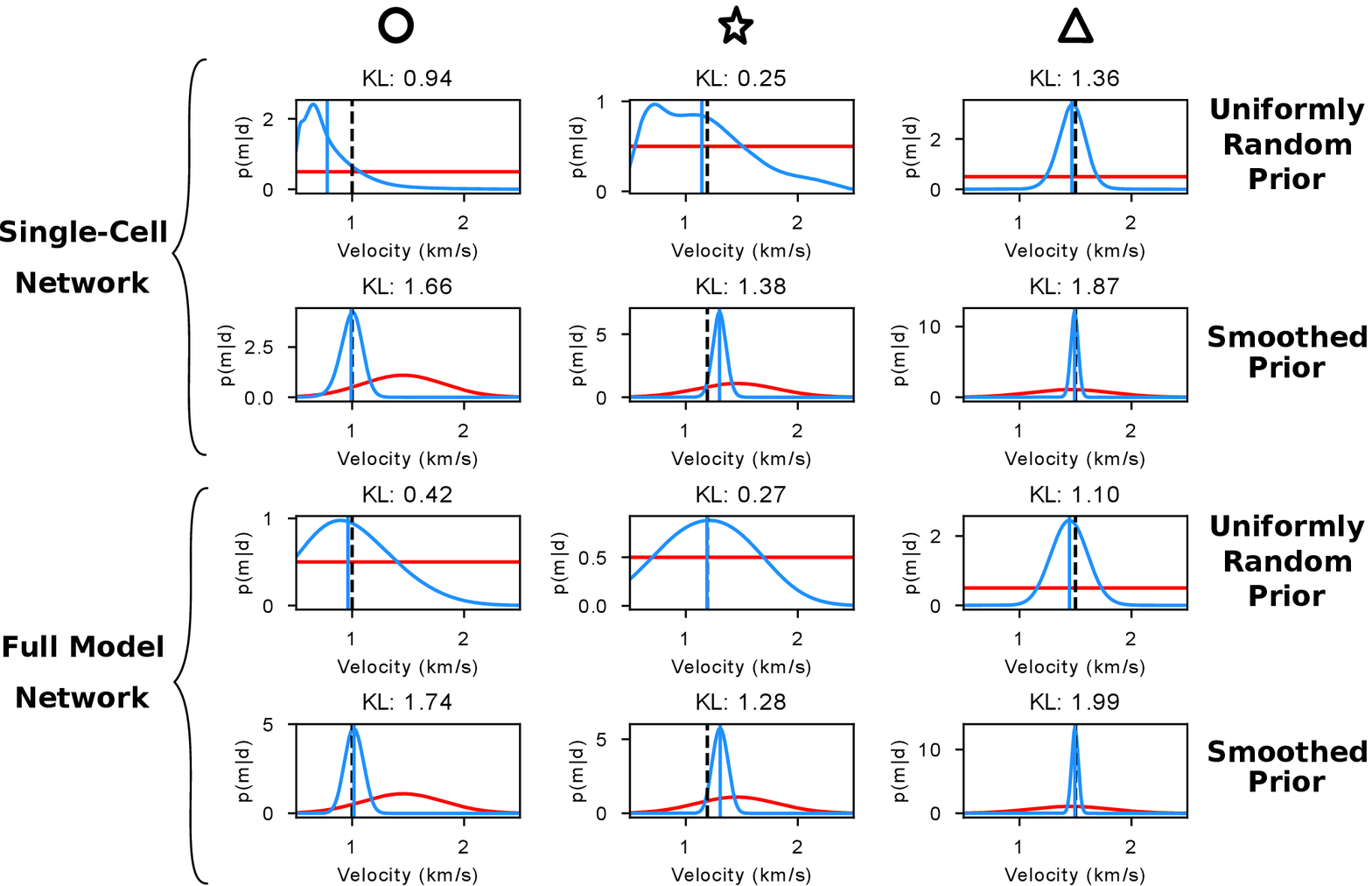}
\caption{Posterior pdfs (blue curves) compared to the prior pdfs (red curves) for the 16x16 grid models for three locations shown in the true model of Figure \ref{fig:model_uncer}: circle (left), star (middle), triangle (right). The rows show results from: (Row 1) Separate-cell MDN's using Uniformly random training dataset. (Row 2) Separate-cell MDN's using the smoothed training dataset. (Row 3) Full-model MDN using Uniformly random training dataset. (Row 4) Full-model MDN using the smoothed training dataset. The mean of the posterior is shown by the blue solid line and the true velocity value by a black dashed line. Corresponding KL divergence values are shown above each result.}
\label{fig:pdfs_uncer}
\end{figure}

The separate-cell networks (one network trained for each cell in the velocity model) allow us to estimate the full marginal posterior probabilities for all cells in the model, and these posterior distributions show how the network represents uncertainty. We show the pdfs for 3 points in the model: inside the velocity anomaly (star), at the edge of the anomaly (triangle), and in the background velocity (circle), where the locations are shown in Figure \ref{fig:model_uncer}a. We can see for the 8 x 8 model using the Uniformly random training set (Figure \ref{fig:pdfs_uncer}a and c) the posterior pdf at the edge of the anomaly has a larger uncertainty indicating that the range of possible velocities spans the velocity of the anomaly and that of the background velocity. This is expected at the edge of an anomaly, the boundaries of which are uncertain: the cells could either be inside or outside of the anomaly, and could therefore assume values of the anomaly (low velocity) or the background model (high velocity). This is the maximum range of velocities expected across the model, hence the largest uncertainties should be around anomaly edges \citep{Galetti2015}.

We do not see uncertainty loops in any model trained on the smoothed models. This makes sense because by imposing prior information that the model is relatively smooth we have removed the possibility to include the effect of spatially sharp contrasts between anomalies and the background velocity model, precluding the types of physical trade-offs that create uncertainty loops. This is represented in the pdfs (Figure \ref{fig:pdfs_uncer}b and d) where the uncertainty is much smaller than in (a) and (e) and where there is no noticeable increase in uncertainties at the boundary of the anomaly. Note that there is again a larger information gain for the results from the smooth training set as shown by the KL divergence values.

\subsection{Realistic Velocity Models}\label{realistic}
\begin{figure}
\centering
\includegraphics[width=\textwidth]{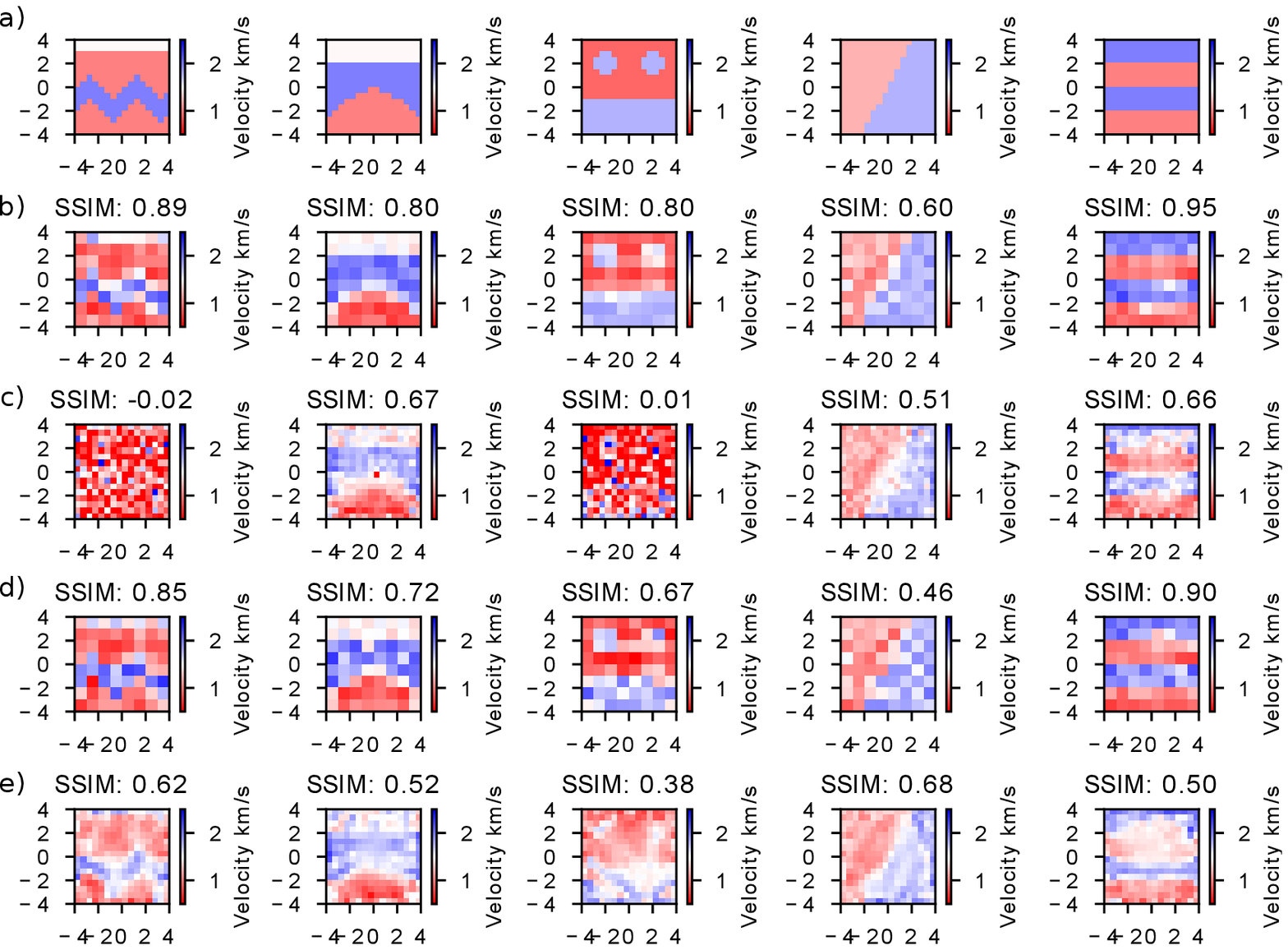}
\caption{(a) True velocity models. Using a random generated training set from a Uniform distribution, mean velocities from separate-cell MDN inversions for (b) an 8 x 8 model and (c) a 16 x 16 model, and from full-model MDN inversion for (d) an 8 x 8 model and (e) a 16 x 16 model. The corresponding SSIM values are shown above each result (see Appendix B for definition of SSIM).}
\label{fig:model_real_random}
\end{figure}

\begin{figure}
\centering
\includegraphics[width=\textwidth]{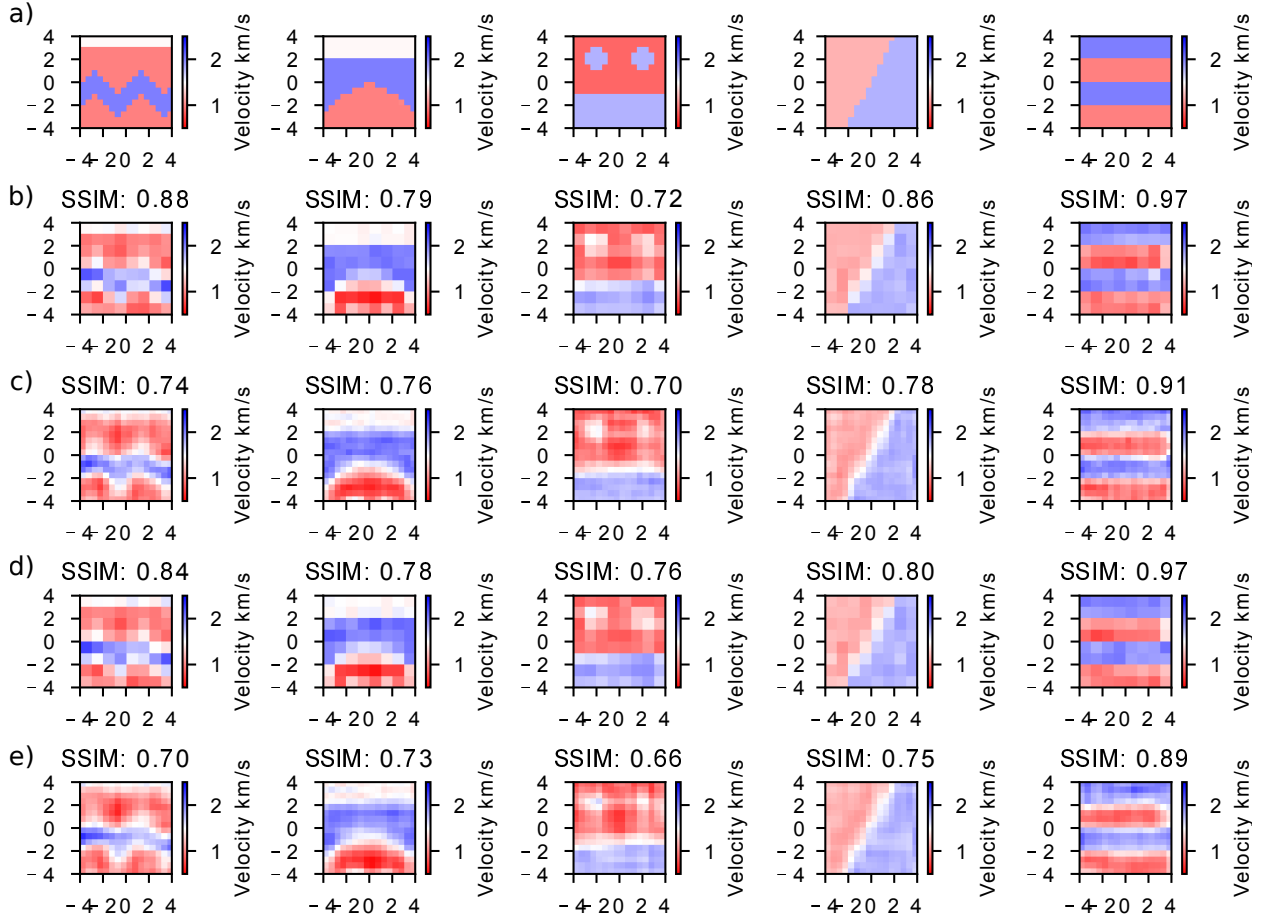}
\caption{(a) True velocity models. Using a  training set drawn of smoothed random models, mean velocities from separate-cell MDN inversions for (b) an 8 x 8 model and (c) a 16 x 16 model, and from full-model MDN inversions for (d) an 8 x 8 model and (e) a 16 x 16 model. The corresponding SSIM values are shown above each result (see Appendix B for definition of SSIM).}
\label{fig:model_real_smo}
\end{figure}

\begin{figure}
\centering
\includegraphics[width=0.6\textwidth]{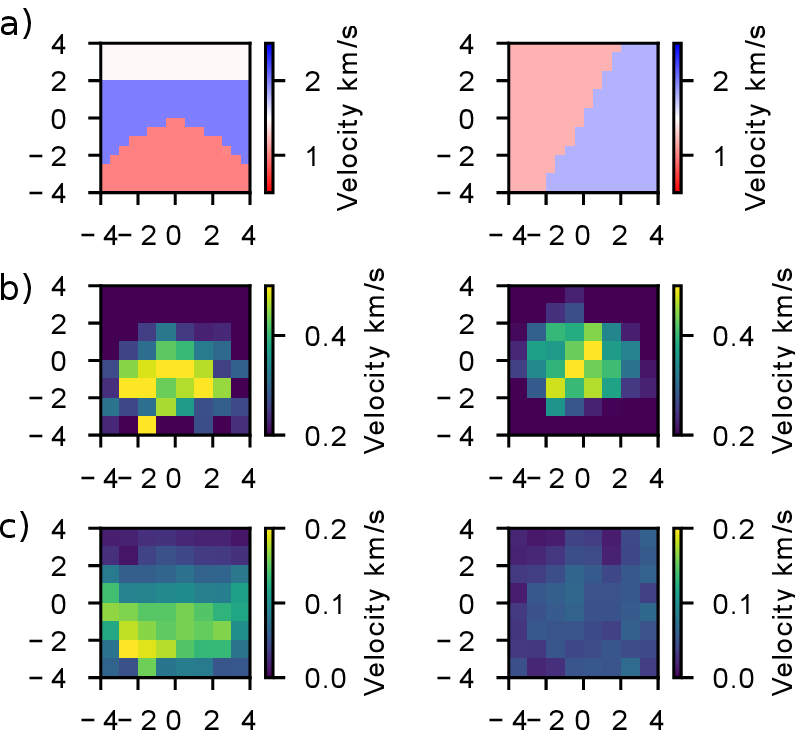}
\caption{(a) True velocity models. For a separate-cell MDN, (b) the standard deviations using a generated training set from a Uniformly random distribution. Using the same network with a training set of smoothed velocities we obtain standard deviations (c).}
\label{fig:std_real}
\end{figure}

Figures \ref{fig:model_real_random} and \ref{fig:model_real_smo} show the results when applying the trained networks to other types of structures that might be encountered in geophysical or non-destructive testing applications. Figure \ref{fig:model_real_random} shows results using Uniformly random training set, whereas Figure \ref{fig:model_real_smo} shows the equivalent results obtained using the smoothed training set. The models inverted on a coarser grid produce reasonable estimates of the velocity models using either prior pdf, however for the smoothed prior all the models, regardless of grid size, are recovered fairly well. Figure \ref{fig:std_real} shows the uncertainty maps for a coarse grid model trained using both types of prior information and inverted using the separate-cell MDN models. When inverting the models with a Uniformly random prior (Figure \ref{fig:std_real}b) the uncertainty maps show a higher uncertainty at the anomaly interfaces (as expected by analogy with the uncertainty loops above), thus helping to define uncertainty in the model geometry, whilst the results from the smooth prior miss this extra information. 

\section{DISCUSSION}\label{discussion}
We compared different methods of mixture density network inversions to estimate tomographic posterior probability density functions. When using datasets with little prior information (Figure \ref{fig:models}a and \ref{fig:models}b) the networks struggle to estimate more than the simplest of velocity models: due to the curse of dimensionality it is simply not possible to provide a sufficient density of prior samples on which to train the MDN. Including stronger prior information in our examples by training on smoothed velocity models (Figure \ref{fig:models}d and \ref{fig:models}e) improves inversion results, although the networks are no longer able to image sharp velocity contrasts, nor estimate uncertainty in the shapes and locations of spatially sharp velocity anomalies, as information about such models is not contained in the training set. Our tests indicate that the prior pdf is the most important factor in improving a network performance since it restricts both the training set and inversion results to a more constrained (effectively lower-dimensional) manifold embedded within the high-dimensional parameter space. This manifold is more densely sampled than the full space thus improving network training and performance. All test models inverted using the stronger prior information give higher SSIM and KL divergence values compared to those using weaker priors, regardless of grid size or how many pixels were inverted with each network. Also, the two circular anomalies in Figure \ref{fig:model_smo} are symmetrical and this symmetry is also shown in all of the smooth-prior inversion results which is not seen in the Uniform-prior results in Figure \ref{fig:model_rand}. Nevertheless, we show that when imposed prior information is false (if the true model is rough but the prior precludes such models) then uncertainty results will be compromised as in Figure \ref{fig:model_uncer}g and \ref{fig:model_uncer}i. It should be noted that neither of the training sets created in this study are fully representative of the true Earth. In reality actual geophyisical features of the Earth are neither uniformly random or smoothed, but dependant on geological charateristics that can include smooth variations or shar boundaries. the results in Figures \ref{fig:model_real_random} and \ref{fig:model_real_smo} show that different structures not seen in the training set can be recovered using this neural network method. However, a clearly advantageous strategy for the future of neural network tomography is to invest effort in finding and using more sophisticated, and correct prior information \citep{Curtis2004}. Recent efforts in this direction include \cite{Walker2014b} who use expert elicitation to constrain prior multi-point geostatistics, \cite{Mosser2018} who use neural networks to parametrise geological prior information, and \cite{Nawaz2017,Nawaz2018,Nawaz2019} who use Markovian models and variational methods with embedded neural and mixture density networks to combine geological and geophysical information; these various directions appear to be strategically important for the future of this field.

\begin{figure}
\centering
\includegraphics[width=0.7\textwidth]{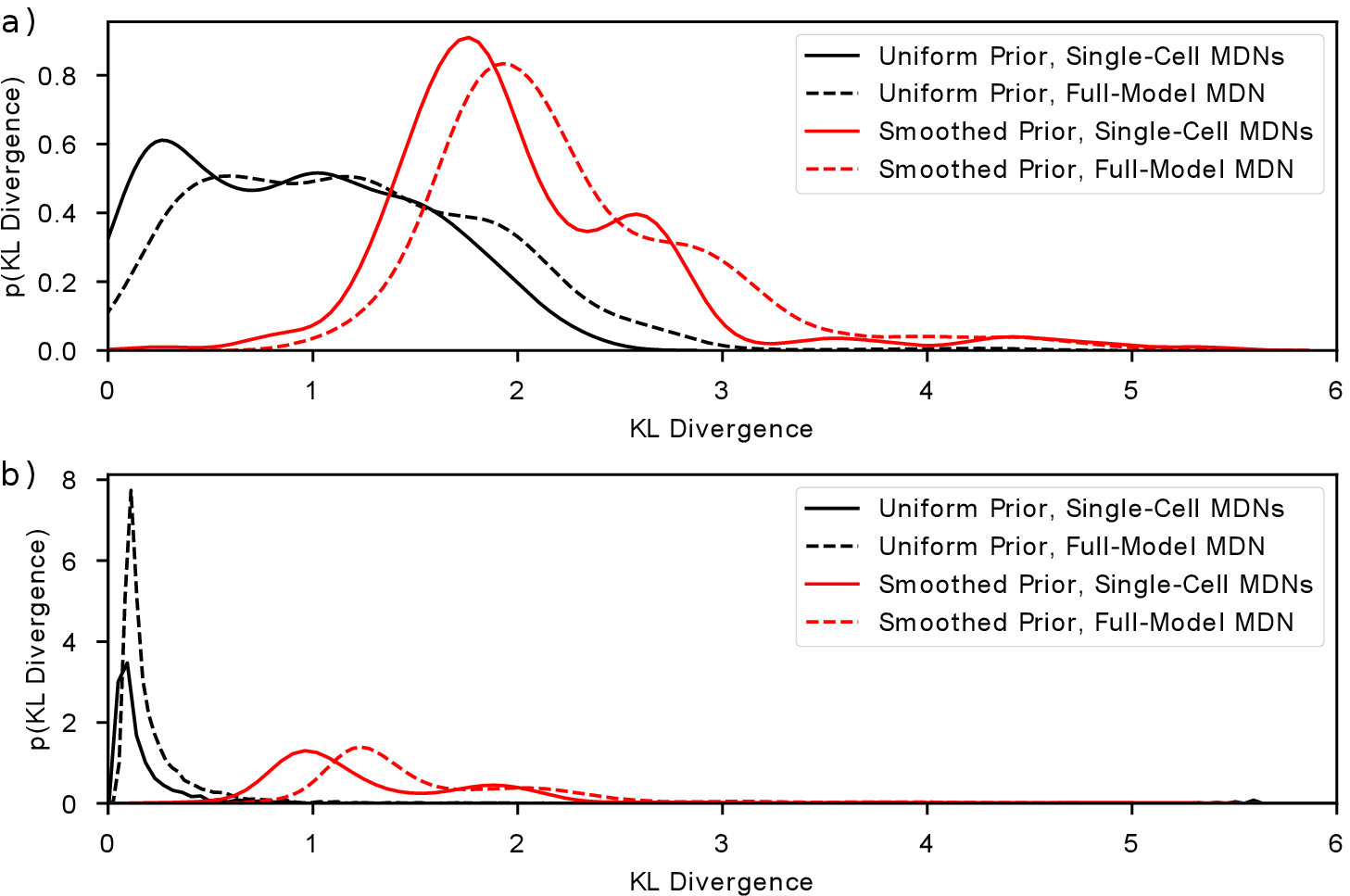}
\caption{Histograms of KL divergence values for results of inverting synthetic data for all models in the test set. a) 8 x 8 models, b) 16 x 16 models.}
\label{fig:KL_div2}
\end{figure}

We illustrate the differences in the KL divergence values in Figure \ref{fig:KL_div2}. The top graph shows histograms of KL values obtained when networks are applied to all synthetic test data for the four different prior and network training types for the 8 x 8 grid model, and the bottom graph is similar but for 16 x 16 models. Both plots confirm that training with a stronger prior increases the information gain in the posterior as was indicated in Figure \ref{fig:pdfs_geol1}. Notice that this is not necessarily an intuitively obvious result: if prior information is weaker or less informative, we might expect the data to add relatively more information, compared to the case where prior information is stronger. We therefore suspect that this result indicates that we simply can not train the MDN's in the case of weaker prior information and sparser training examples; even though by adding stronger prior information we should \textit{decrease} the relative value of the data, this effect is out-weighed by the fact that we can better train the network and thereby extract \textit{more} information from data.

The effect of increasing the number of cells in the model is also clearly highlighted: Figure \ref{fig:KL_div2}a has higher KL values than Figure \ref{fig:KL_div2}b. Interestingly, both plots show that training using a full-model inversion slightly increases the KL divergence, implying that the networks are making use of the relationship between adjacent pixels to better constrain the posterior pdfs.
\subsection{Inference limits}\label{fig:inv_limits}
\begin{figure}
\centering
\includegraphics[width=\textwidth]{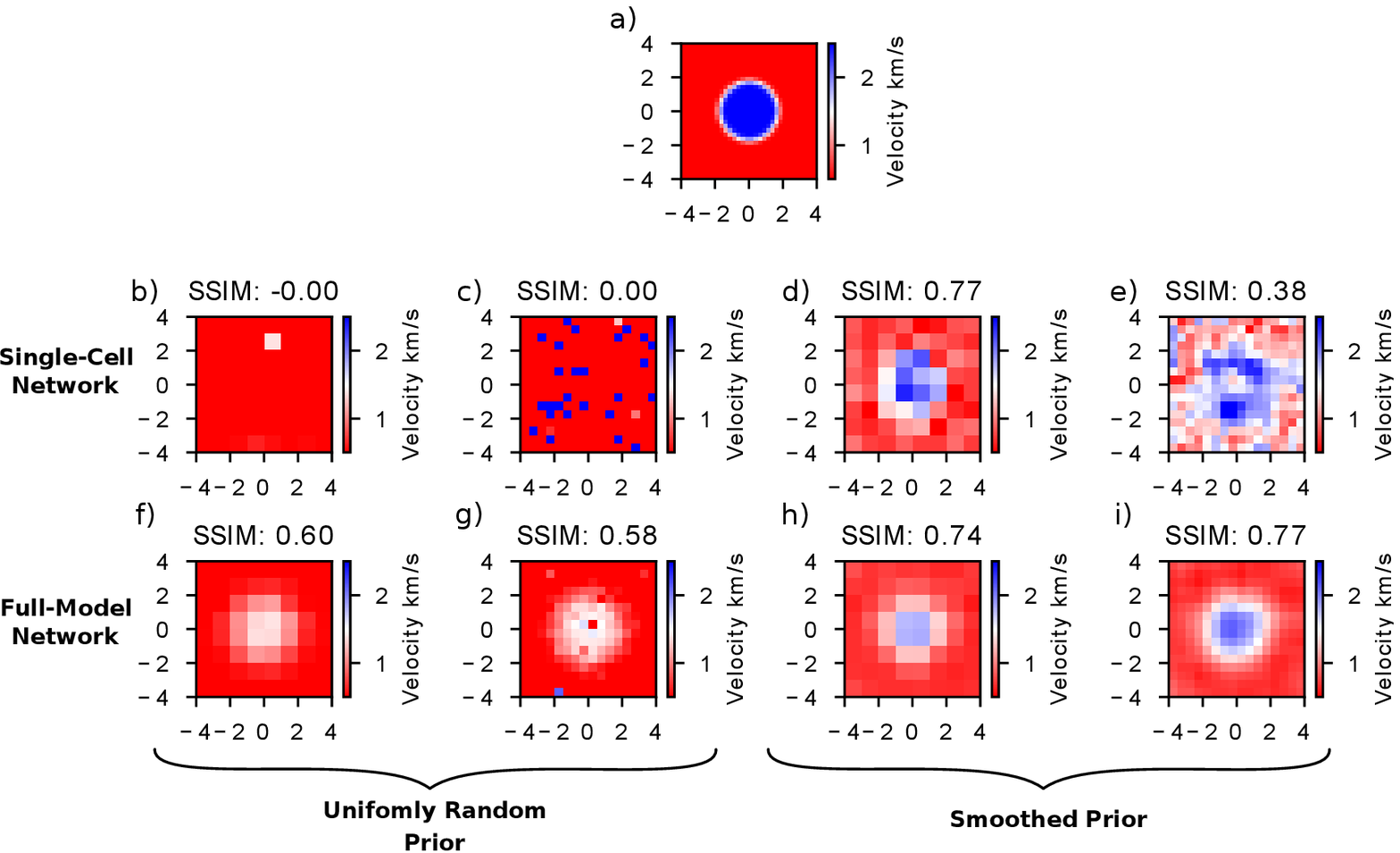}
\caption{(a) True velocity model. For a separate-cell MDN, using a training set from a Uniformly random distribution, results shown are mean velocities for (b) an 8 x 8 model and (c) a 16 x 16 model. Using the same network with a training set of spatially smoothed velocities we obtain mean velocities for (d) an 8 x 8 model and (e) a 16 x 16 model.  For a full-model MDN, using a training set from a Uniformly random distribution we obtain mean velocities for an (f) 8 x 8 model and (g) a 16 x 16 model. Using the same network with a training set of spatially smoothed velocities we obtain mean velocities for an (h) 8 x 8 model and (i) a 16 x 16 model. Corresponding SSIM is shown above each result. The colour axis has been clipped to the velocity bounds of the training set $(0.5km/s,2.5km/s)$.}
\label{fig:model_lim}
\end{figure}

When creating the training dataset we set hard bounds on the grid cell velocities, thus limiting the range of velocity models that should be found using the trained networks. Figure \ref{fig:model_lim} shows the inversion of a model at the limits of all training datasets. The middle row shows results when using the Uniformly random training dataset: none of the inversions give reliable results. Although the network trained to invert the full model at once performs slightly better, all networks produce extremely poor results. This is expected as the velocity model has a background velocity at the lower limit of the training sets, $0.5km/s$ and an anomaly at the upper limit, $2.5km/s$. This is an extreme example that is not likely to have proximal samples in the training set, therefore the results are expected to be poor.

The same model lies outwith the dataset with a stronger prior as well, but networks appear to recognise that there is a velocity anomaly. However, since the prior dataset used is smoothed, strong contrasts are precluded and none of the networks give accurate velocity information, despite being able to represent the geometry of the structure.

\subsection{Inversion Speed}\label{inv_speed}
As this is a prior sampling method the training dataset must be created in advance. It took  $t_{prior} =$ 11 hours, to create the training dataset of 2.5 million samples using 5 CPUs on a Dell PowerEdge R820. However, this only needs to be done once; even if more prior information becomes available we may be able update our prior using the \textit{prior replacement} method of \cite{Walker2014} or the resampling method of \cite{Sambridge1999} rather than calculating entirely new training examples.

In this work each network took between 1-2 hours to train (converge) using 16GB of RAM over 2 NVIDIA TITAN X GPUs. For the 8 x 8 grid models with an ensemble of 8 networks when training the network for each grid cell separately, we required $8 \times 8 \times 8 = 512$ networks in total and for the 16 x 16 models with an ensemble of 4 networks we required $16 \times 16 \times 4 = 2048$ networks. However, the training of each network is independent of others so the process can easily be parallelised and using 50 cores a full training run for the larger 16 x 16 grid model takes $t_{train} = $ 80 hours of real clock time. For the full-model networks only one network is trained for all cells so the total training time is much lower: each network takes around 3 hours to train using 48GB of RAM oer 4 NVIDIA TITAN X GPUs so training 10 networks only takes 30 hours without running them in parallel. This process could be reduced to 3 hours by using only 10 cores and reduced further by training each network across cores. The advantage of an MDN is the speed of inversion after training: once a network is trained new inversions take a fraction of a second, even on a standard desktop computer. Computational efficiency is therefore gained only when the trained networks will be applied to many different data sets.

Monte Carlo methods are known to be computationally expensive \citep{Bodin2009} and a fully non-linear Markov chain Monte Carlo (McMC) tomographic inversion can take weeks or months of compute time. Monte Carlo methods use posterior sampling so for every new inversion a new sample set must be performed. This is often a far less demanding sampling task than sampling with similar density of samples from the prior since high probability parts of the posterior pdf usually span a significantly smaller volume of parameter space. Nevertheless, neural network methods are advantageous over traditional Monte Carlo methods when $n$ repeated inversions of similar data types are to be performed provided that $n > \frac{(t_{prior}+t_{train})}{t_{MC}}$, as the computationally expensive sampling step only needs to be performed once and the network-based inversion becomes faster. In a tomographic setting this could be useful for monitoring purposes, where data collected periodically from the same set of sources and receivers can be inverted with the same network(s) each time new data arrives. However it should be noted that despite the longer computation time Monte Carlo methods can be used to produce higher resolution 2D or 3D models \citep{Galetti2016,Zhang2019}. Mixture density networks have also been shown to give conservative uncertainty estimates compared to Monte Carlo methods \citep{Kaufl2016}.
\subsection{Training Flexibility}\label{flex}
In this work we train networks assuming that the data (travel times) are recorded with exactly the same data acquisition geometry as was used for training. It would also be possible to train more flexible networks that account for missing data. For example, one could augment the training set with additional samples constructed from the same data-model pairs ${(\mathbf{d}_i,\mathbf{m}_i): i =1,...,N}$ but with a certain number of travel time values in the dataset randomly set to $0$, to indicate a missing value \citep{DeWit2013}. Then new datasets with a missing values (for example due to a noisy stations causing errors in travel times) can be inverted by the same network.

Data from new receivers added after training the network cannot be used. However we can create a new training set containing only the data from the added receiver station and fine tune the original network by using the original network parameter values as a starting point for training optimisation. This has the advantage that the training process will be much faster.
\section{CONCLUSION}\label{conclusion}
We present neural network-based, non-linear inversion methods applied to a 2D travel time tomography problem to estimate posterior probability density functions. The flexibility of mixture density networks mean that we can provide uncertainty estimates for 2D velocity maps. We show that the prior information used to create the training dataset is the most important factor in providing accurate velocity estimates and uncertainties as such information effectively reduces the dimensionality of the tomography problem. However, as with all Bayesian inversions if we impose false prior information we can lose important information about uncertainties. By training networks to invert for a full tomographic model at once, we can also understand the relationship between velocities in neighbouring pixels; however the number of parameters in the inversion increases substantially, and training for accurate models proves to be significantly more difficult. We compare the speed of neural network inversion to more standard Monte Carlo methods and determine that for many repeated inversions such as occur in monitoring situations, MDNs may out-perform Monte Carlo methods in terms of computational cost. 

\section*{Acknowledgements}
The authors thank the Edinburgh Interferometry Project sponsors (Equinor, Schlumberger Cambridge Research and Total) for supporting this research.

\section*{Appendix 1: Network configurations}\label{Append1}
The networks trained on individual cells used 4 fully connected layers (FC), where each node receives an input from every node in the previous layer. In between each node of the fully connected (FC) layers a rectified linear unit (ReLU) is used. The individual layer sizes and the total number of parameters to be trained in each networks is outlined in Table \ref{net_config}.
\begin{table}[htbp!]
\centering
\begin{tabular}{cccccc}
\hline
  Size of model & FC 1 & FC 2 & FC 3 & FC 4  		& Total No. of Parameters\\ \hline
   & 270            & 1000           & 380            & 600     & 1,544,765       \\
   & 100            & 500            & 450            & 550      &  1,622,685  \\
   & 800           & 325            & 100            & 300       &  1,165,660   \\
8 x 8  & 200            & 400            & 200            & 50 &  334,335         \\
   & 300            &250            & 200            & 50         &  331,685\\
   & 900            & 700            & 70            & 550        &  2,077,505   \\
   & 200            & 250           & 200           & 50          &  274,185 \\
   & 300           & 400            & 200            & 50         &  406,835  \\ \hline
   & 375            & 500            & 300            & 600      &  5,265,470     \\
16 x 16 &300            & 250            & 200            & 50& 625,445            \\
   & 200           & 400            & 200            & 50         &  628,095  \\
   & 800            & 1000            & 500            & 550    &  6,076,995       \\ \hline
\end{tabular}
\caption{Network configurations of the networks with 4 fully connected (FC) layers. Each row in the table represent a separate networks trained. Eight networks were trained for the 8 x 8 models and four networks for the 16 x 16 models.}
\label{net_config}
\end{table}

Networks trained on the whole model (all cells at once) used a convolutional network with 3 convolutional layers (Conv) and 4 fully connected layers. The sizes of each layer and the total number of parameters to be trained in each networks is outlined in Table \ref{net_config2}.
\begin{table}[htbp!]
\centering
\begin{tabular}{cccccccccccc}
\hline
\multicolumn{2}{c}{Conv 1} & \multicolumn{2}{c}{Conv 2} & \multicolumn{2}{c}{Conv 3} & FC 1 & FC 2 & FC 3 & FC 4 & \multicolumn{2}{c}{Total No. of Parameters}\\ \hline
Filter              & Kernel              & Filter              & Kernel              & Filter              & Kernel              &                &                &                &              &8x8				& 16x16  \\ \hline
128                 & 5                   & 128                 & 5                   & 64                  & 1                   & 800            & 150            & 600            & 1500        & 4,717,405  & 13,363,165	\\
32                  & 9                   & 32                  & 5                   & 16                  & 1                   & 500            & 300            & 600            & 1500          & 4,354,183	  &12,999,943	\\
32                  & 9                   & 32                  & 5                   & 16                  & 1                   & 500            & 200            & 2000           & 1250         & 5,641,438  &12,847,243	\\
32                  & 9                   & 8                   & 5                   & 16                  & 1                   & 500            & 300            & 600            & 1750           & 4,986,575  &15,054,335	\\
32                  & 9                   & 32                  & 5                   & 16                  & 1                   & 500            & 1500           & 50             & 1250          & 3,528,333  &10,734,093	\\ \hline
\end{tabular}
\caption{Network configurations of the convolutional networks with three convolutional (Conv) layers and 4 fully connected (FC) layers. Each row in the table represent a separate networks trained.}
\label{net_config2}
\end{table}

\section*{Appendix 2: Structural Similarity Index Measure (SSIM)}\label{Append2}
We use the form of the SSIM metric described in \cite{Wang2004}. Let $x$ and $y$ be a window of $N$x$N$ size. We calculate the luminance $l(x,y)$, contrast $c(x,y)$ and structure $s(x,y)$ defined as:
\begin{equation}
l(x,y) = \frac{2\mu_x\mu_y + C_1}{\mu^2_x + \mu^2_y + C_1}
\end{equation}
\begin{equation}
c(x,y) = \frac{2\sigma_x\sigma_y + C_2}{\sigma^2_x + \sigma^2_y + C_2}
\end{equation}
\begin{equation}
s(x,y) = \frac{\sigma_{xy} + C_3}{\sigma_x\sigma_y + C_3}
\end{equation}
where $\mu$ and $\sigma$ are the mean and variance of the windows $x$ or $y$ and $\sigma_{xy}$ is the covariance of $x$ and $y$. To avoid instability in the division, constants $C_1$, $C_2$ and $C_3$ are defined as $C_1 = (k_1L)^2$ and $C_2 = (k_2L)^2$ where $L$ is the dynamic range of the cell values while $k_1=0.01$ and $k_2 =0.03$, and $C_3 = C_2/2$. The three components are combined to give the full SSIM:
\begin{equation}
SSIM(x,y) = [l(x,y)^\alpha \cdot c(x,y)^\beta \cdot s(x,y)^\gamma]
\end{equation}
where $\alpha$, $\beta$ and $\gamma$ are weighting parameters. Setting $\alpha=\beta=\gamma=1$ we can simplify the expression to:
\begin{equation}
SSIM(x,y) = \frac{(2\mu_x\mu_y + C_1)(2\sigma_{xy}+C_2)}{(\mu^2_x + \mu^2_y + C_1)(\sigma^2_x + \sigma^2_y + C_2)}
\end{equation}
We perform the calculation over sliding windows and take the mean of the resulting $SSIM(x,y)$ values. For the 8 x 8 models we use 3x3 windows and the 16 x 16 models use 7x7 windows, so that the windows cover a similar spatial area.

\bibliographystyle{natbib}  
\bibliography{TomoNN} 

\end{document}